\documentclass[a4paper,11pt]{article}

\usepackage{jinstpub}

\usepackage{wrapfig}
\usepackage{subcaption}
\usepackage{cleveref}
\usepackage{lineno}

\title{Development of the New Optical Sensor for IceCube-Gen2}

\collaboration[c]{on behalf of the IceCube-Gen2 Collaboration$^*$\note[*]{Full author list and acknowledgments are available at \href{https://icecube.wisc.edu/collaboration/authors/\#collab=IceCube&date=2025-08-25&formatting=web}{icecube.wisc.edu}.}}

\author{D. Butterfield and C. Wendt}
\affiliation{University of Wisconsin - Madison,\\Madison, WI 53706, USA}

\emailAdd{dbutterfield@icecube.wisc.edu}
\emailAdd{chwendt@icecube.wisc.edu}

\abstract{A new digital optical module (DOM) has been developed for the proposed expansion to the IceCube detector at the South Pole, IceCube-Gen2. The “Gen2-DOM” has 4 times the integrated photon sensitivity of the current IceCube DOMs and has built off the design features of the IceCube Upgrade modules. The Gen2-DOM has up to 18 4" photomultiplier tubes (PMTs) in a borosilicate glass pressure vessel, arranged in a uniform 4$\pi$ angular distribution. The mechanical design has been optimized to fit into a reduced borehole diameter which, in turn, will reduce drilling costs during installation. Each PMT has a dedicated readout board, designed to increase sensitivity to high-energy events aligned with the science goals of IceCube-Gen2. Internal storage enables multi-level triggering schemes with reduced overall flow of data on the long cables. Twelve prototypes of the Gen2-DOM will be deployed in the IceCube Upgrade in the 2025/2026 austral summer. This article will focus on the current status of design development and initial performance testing results.}

\proceeding{Presented at the 32nd International Symposium on Lepton Photon Interactions at High Energies\\
  Madison, Wisconsin, USA\\
  25-29 August, 2025}

\notoc
\begin{document}
\maketitle
\flushbottom

\vspace{-1em}
\section{Introduction}
\label{sec:introduction}

IceCube-Gen2, the proposed expansion to the IceCube South Pole Neutrino Observatory \cite{IceCube_Detector}, plans to enhance the detection capabilities for astrophysical neutrinos. The 8~km$^3$ in-ice optical array will be composed of 120 strings, with 80 optical modules per string and a total of 9600 optical modules. With the science goals focused on high energies \cite{Gen2_whitepaper}, the array will be sparser than IceCube, with inter-string spacing of 240 m instead of 125 m. With such a sparse array, the optical modules must have high photon detection efficiency and dynamic range, intended to achieve a factor of four improvement compared to the existing Digital Optical Modules (DOMs) in IceCube. In addition, the newly designed Gen2-DOMs must fit within a power budget of < 4 Watts / module, and have a diameter of less than 12.5 inches to save drilling costs during installation.

\vspace{-1em}
\section{Sensor Design}
\label{sec:sensor_design}

\begin{figure}[htbp]
\centering
\includegraphics[width=.3\textwidth]{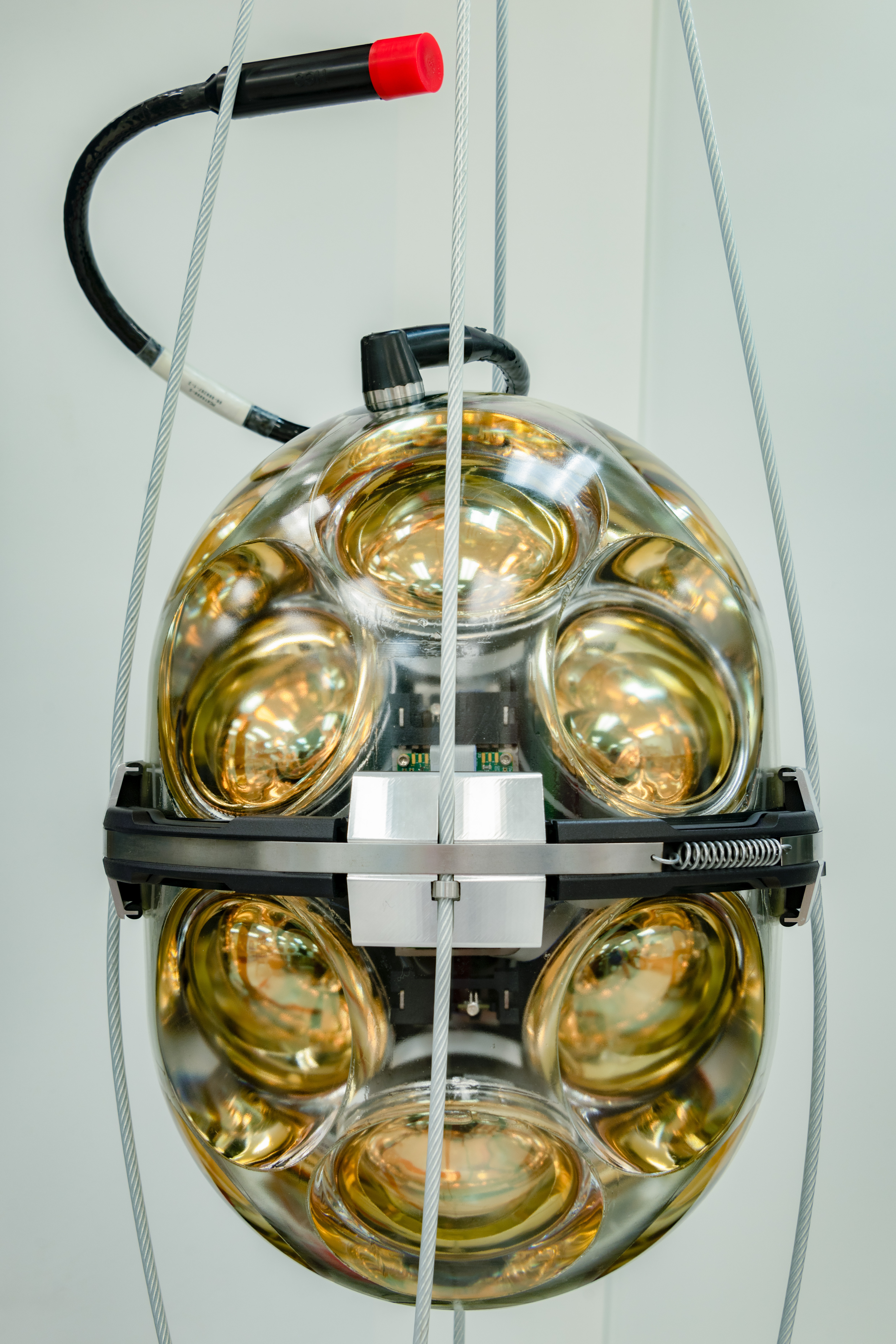}
\includegraphics[width=.3\textwidth]{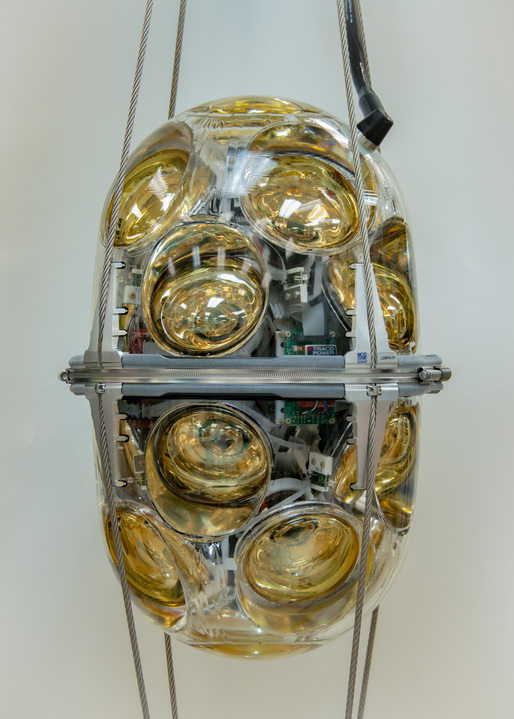}
\caption{\label{fig:gen2doms} Completed prototype for Gen2 Design Candidate-16 (left) and  Gen2 Design Candidate-18 (right).}
\end{figure}

Building off the reliability of the IceCube DOM and the design features of the mDOM \cite{mdom} and DEgg \cite{degg} modules of the IceCube Upgrade \cite{upgrade}, the Gen2-DOM has a segmented photocathode surface with up to 18 photomultiplier tubes (PMTs) arranged with uniform 4$\pi$ angular coverage inside an elongated glass pressure vessel. The design of the new sensors will be tested in-situ in the IceCube Upgrade, where 12 prototype modules will be deployed in the 2025/26 austral summer. Two prototype designs of the Gen2-DOM have been implemented: one with 16 PMTs and one with 18 PMTs. They are called the "Gen2 Design Candidate-16" (Gen2DC-16) and "Gen2 Design Candidate-18" (Gen2DC-18), respectively, and the completed designs are shown in Figure~\ref{fig:gen2doms}. Both prototype designs have the same central components, while testing different integration methods and component manufacturers. Six of each prototype will be deployed in the IceCube Upgrade, and the different designs will be merged into one module which is developed and optimized for mass production: the Gen2-DOM. The primary components of each of the prototypes are discussed in the following sections.

\vspace{-0.5em}
\subsection{Photomultiplier Tubes}
\label{subsec:pmts}

Both the number of PMTs in the module and the size of each PMT were considered when designing each of the Gen2-Design Candidates. Studies to optimize the effective area of the modules showed a 4" diameter PMT with a short neck (105mm total length) as the best option. This study was heavily informed by the 3" PMT design used in the mDOMs. We approached two manufacturers to design these PMTs: Hamamatsu Photonics KK (HPK) \footnote[1]{Hamamatsu Photonics K.K., Shizuoka, Japan} and North Night Vision Technology (NNVT) \footnote[2]{North Night Vision Technology Co., Ltd, Nanjing, China}. Both companies proposed their initial designs for new 4" PMT models, which were then characterized thoroughly and met all requirements \cite{Gen2_PMTs}. With almost 10,000 modules planned for IceCube-Gen2, having two PMT manufacturers benefits the rate and security of mass production.

\vspace{-0.5em}
\subsection{Pressure Vessel}
\label{subsec:pressure_vessel}

The two modules' designs have different number of PMTs, and therefore have different shaped pressure vessels. Two manufacturers were chosen for the different modules. The Gen2DC-16 vessel is produced by Nautilus \footnote[3]{Vitrovex by Nautilus Marine Service, GmbH, Buxtehude, Germany} and has an outer diameter of 312 mm and a height of 444 mm. The Gen2DC-18 vessel is produced by Okamoto \footnote[4]{Okamoto Glass Co., Ltd, Kashiwa, Chiba, Japan} and has a more elongated shape, with a diameter of 318 mm and height of 540 mm. Each vessel is designed to protect the interior components of the module, which faces pressures up to 550 bar as measured during IceCube deployment. Both vessels are rated up to 700 bar, have diameters less than 12.5 inches, low levels of radioactive contamination, and good transmissivity between 300-500 nm. Both vessels are made of borosilicate glass with thickness in the range of 12-16.5 mm and meet all design requirements.

\vspace{-0.5em}
\subsection{Gel Pads}
\label{subsec:gelpads}

Any air gaps between the photocathode and the pressure vessel can cause photon losses due to the differences in index of refraction. To minimize these losses, the PMTs are coupled to the vessel with silicone optical gel. The index of refraction of the gel at 400 nm is 1.41, which is similar to that of borosilicate glass, 1.49. The gel pad acts as a sort of funnel for the photons, guiding them to the photocathode via total internal reflection.

We cast these gel pads in a two step process. The first step is to cast a hollow gel pad directly onto the photocathode face with a custom mold that creates a highly smooth exterior with a ring shaped boundary \cite{Gen2DOM_mechanical}. We then adhere the edges of the hollow gel pad to the interior of the pressure vessel with gel, and finally backfill the hollow cavity with uncured gel. The gel then cures directly to the glass, creating a strong and bubble-free bond.

\vspace{-0.5em}
\subsection{Electronics}
\label{subsec:electronics}

The Gen2-DOM utilizes newly developed PMT readout bases with self-contained DAQ functionality, called the wuBase \cite{Gen2DOM_electronics}. The wuBase includes the same active Cockcroft-Walton high voltage generator developed for the IceCube Upgrade's mDOM PMTs, with the new DAQ functions on an extended part of the board which fits into spaces between PMTs in the Gen2-DOM. Each base includes a fully digitized readout with a two channel, 12-bit ADC at a rate of 60 megasamples per second (MSPS) and a single photoelectron (SPE) timing resolution of 2.5 ns. The ADC digital outputs are streamed to a low power FPGA where triggered waveforms are accumulated in buffers.  The wuBase includes an ARM microcontroller that reads these waveforms, performing first level computations as well as managing all other DAQ and control functions. The high gain channel records the waveform from the anode, while the low gain channel records the signal from dynode 8 out of 10. This combination achieves the high dynamic range of 5000 photoelectrons within 25 ns, with electronic noise below 2$\%$ of the SPE level.

The self-contained functionality of the wuBases goes along with the use of a "Mini Mainboard" (MMB), which serves as the command and communications hub for the Gen2-DOM, while squeezing into a space between PMTs. Two fanout boards have been developed to multiplex the communication between the MMB and the wuBases, which operates at 3 Mbaud.

\vspace{-1em}
\section{First Lab Measurements}
\label{sec:lab_measurements}

\begin{figure}[htbp]
     \begin{subfigure}[t]{0.3\textwidth}
        \centering
        \includegraphics[width=\textwidth]{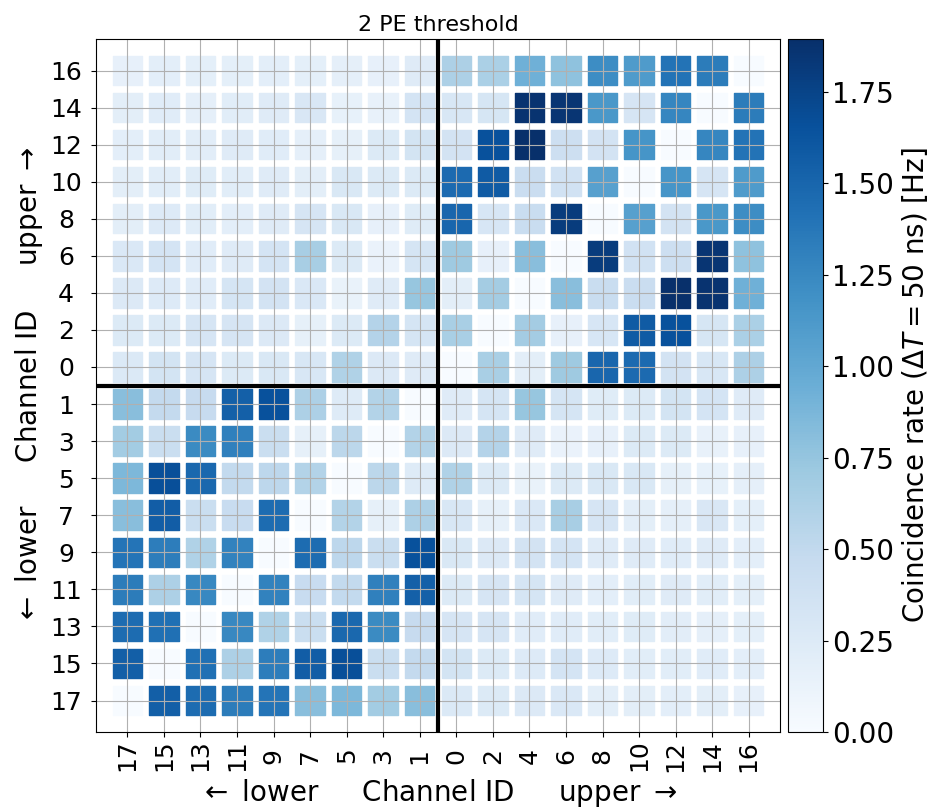}
        \label{fig:coinc_2pe}
        \caption{\label{fig:coinc_2pe} Dark noise coincidence rates on pairs of PMTs for 2 PE threshold.}
     \end{subfigure}
     \hfill
     \begin{subfigure}[t]{0.3\textwidth}
        \centering
        \includegraphics[width=\textwidth]{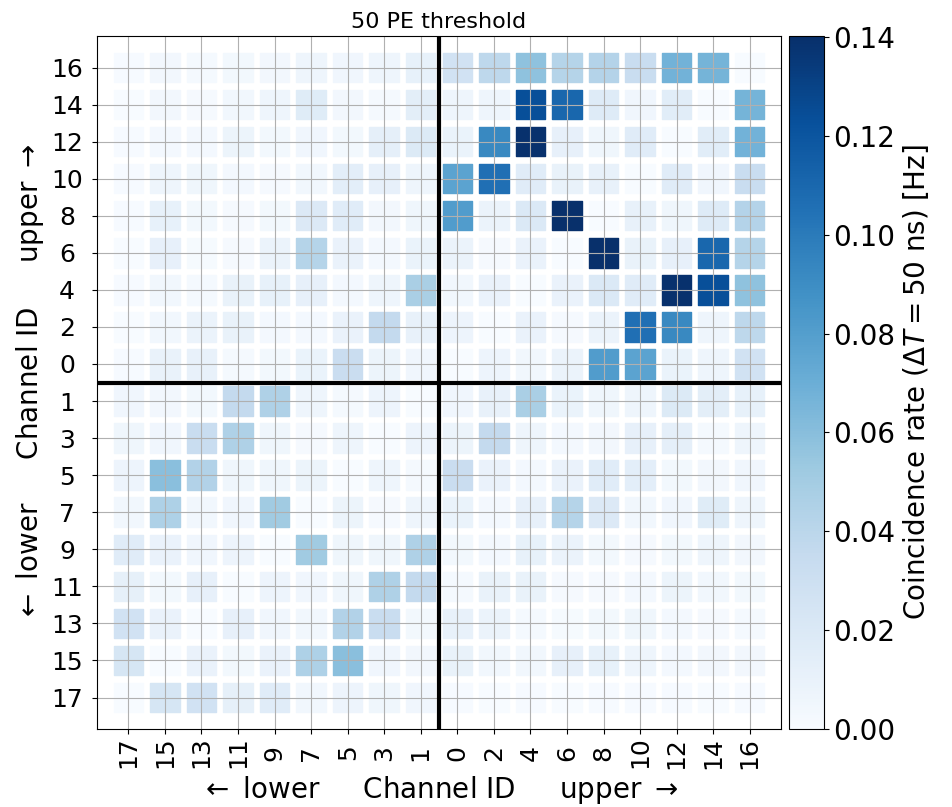}
        \label{fig:coinc_50pe}
        \caption{\label{fig:coinc_50pe} Dark noise coincidence rates on pairs of PMTs for 50 PE threshold.}
     \end{subfigure}
     \hfill
     \begin{subfigure}[t]{0.3\textwidth}
        \centering
        \includegraphics[width=\textwidth]{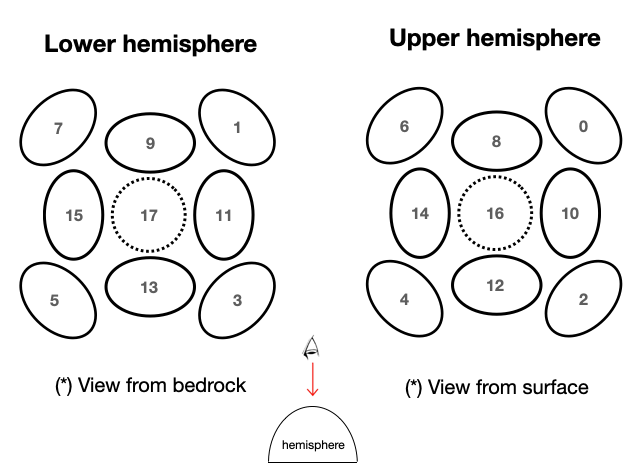}
        \label{fig:pmt_nums}
        \caption{\label{fig:pmt_nums} Channel ID number assignment in Gen2 Design Candidates.}
     \end{subfigure}
     \label{fig:coinc_plots}
     \caption{\label{fig:coinc_plots}Dark noise coincidence rates and PMT numbering scheme.}
\end{figure}

Before integration into the full module, each PMT was tested as a subassembly with its wuBase and gel pad already attached. The tests were performed at low temperatures, including PMT gain, electronics performance, dark count rate, and time resolution \cite{Gen2DOM_overview}.  The pre-integration dark rates for the PMTs are fairly low, between 100-250 Hz. The dark noise of the fully integrated module is dominated by radioactive contaminants in the pressure vessel. The radioactive decay processes produce bursts of scintillation photons, which can be captured across multiple PMTs. Figure~\ref{fig:coinc_2pe} showcases the coincidence rate between pairs of PMTs within a time window of 50 ns and at a threshold of 2 PE. The PMT numbering scheme is shown in Figure~\ref{fig:pmt_nums}. We can see there is a strong symmetry between the hemispheres, with neighboring PMTs showing the highest coincident noise rates. If we increase the threshold to 50 PE, the coincidence rate of noise hits is largely suppressed. In Figure~\ref{fig:coinc_50pe}, we see that high-charge coincidences are observed much more in the upper hemisphere. The pattern and rate support interpretation of these events as down-going atmospheric muons.

\vspace{-1em}
\section{In-Ice Triggering Scheme}
\label{sec:trigger}

\begin{figure}[htbp]
\centering
\includegraphics[width=.45\textwidth]{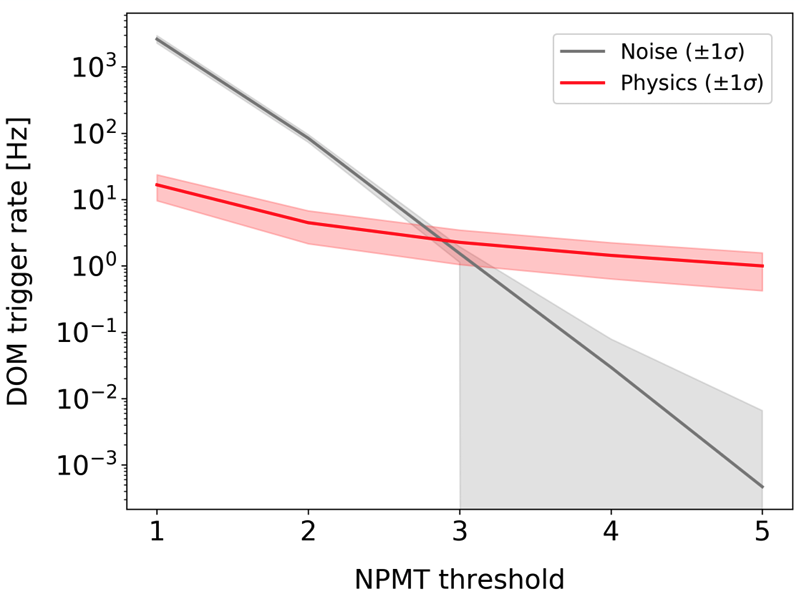}
\caption{\label{fig:gen2_trig}Simulated cosmic rays trigger more than noise hits when requiring a minimum of 3 PMT channels to record a hit.}
\end{figure}

IceCube-Gen2 will use a multi-level trigger and readout protocol, which will reduce the data flow on the long cables to below the level of IceCube. In the new trigger scheme, PMT hit details are temporarily saved in in-module flash memory buffers. If a number-of-PMTs threshold is met in the module, a trigger message is sent to the surface. At the surface, streams of trigger messages are combined while selecting events with defined requirements on numbers of modules and strings contributing in a time window. Only for these time windows, the system will issue readout requests for all corresponding stored hits, which greatly suppresses the amount of noise hits sent to the surface. A detailed simulation with various candidate requirements shows that physics hits trigger the module more often than noise hits when requiring a minimum of 3 (out of 16 or 18) PMTs to record a hit within a 500 ns window \cite{tdr}, as shown in Figure~\ref{fig:gen2_trig}. 

In reducing the bandwidth usage up the main cable assembly, IceCube-Gen2 is able to increase the number of devices per wire-pair on the cable. In IceCube, there are 2 DOMs per wire-pair; in the Upgrade there are 3 devices per wire-pair; in IceCube-Gen2 there will be 6 devices per wire-pair. Taking into consideration the increase in photon sensitivity  per device, this results in a factor of 18 increase in photon detection efficiency per wire-pair, and a substantial savings in cost and logistics from reduced cable size.

\vspace{-1em}
\section{Conclusion and Outlook}
\label{sec:conclusion}

In order to meet the design and functionality requirements of IceCube-Gen2, two sensor designs have been developed in parallel to test the practical aspects of electronic and mechanical assembly, as well as performance in the ice. While both designs feature the same central components of PMTs, electronics, and gel pads, each design offers a beneficial internal structure concept and assembly processes which will contribute to the final Gen2-DOM. Both the Gen2DC-16 and Gen2DC-18 fulfill all requirements and six prototypes of each module will be tested in-situ as part of the IceCube Upgrade. Work to merge the two designs into a final Gen2-DOM design is underway, with the focus lying on manufacturability and reliability for mass production.

\bibliographystyle{JHEP}
\bibliography{main}

\end{document}